\documentclass[conference]{IEEEtran}
\usepackage{amsmath}
\usepackage{amsthm}
\usepackage{amssymb}
\usepackage{graphicx}
\usepackage{algorithm}
\usepackage{algpseudocode}
\usepackage{hyperref}
\usepackage{cleveref}

\title{Probabilistic Trust-Based Enhancement for simultaneous transmission in AOMDV Routing Protocol}

\author{
    \IEEEauthorblockN{Nikhil Mishra}
    \IEEEauthorblockA{
    Department of Electrical Engineering\\
    Indian Institute of Technology, Kanpur\\
    \href{mailto:mnikhil21@iitk.ac.in}{mnikhil21@iitk.ac.in}}
}

\begin{document}

\maketitle

\begin{abstract}
This work addresses a trust-based enhancement to the Multipath Ad hoc On-Demand Distance Vector (AOMDV) routing protocol. While AODV and its multipath variant AOMDV have been fundamental in mobile ad hoc networks, they lack mechanisms to account for node reliability. A probabilistic link-trust model is proposed that incorporates factors such as past behavior, battery levels, and node coupling to distribute data optimally to reduce delay while simultaneously transmitting through multiple paths.
\end{abstract}

\section{Introduction}
Mobile Ad hoc Networks (MANETs) present unique challenges in routing due to their dynamic topology and decentralized nature. These networks must adapt to frequent changes in connectivity, varying node capabilities, and potential node failures. The Ad hoc On-Demand Distance Vector (AODV) protocol and its multipath variant AOMDV have been widely adopted for their simplicity and effectiveness. However, these protocols do not account for node reliability.

Traditional routing protocols often make decisions based solely on hop count or basic metrics, neglecting factors such as historical node reliability and energy constraints. This limitation can lead to suboptimal path selection and unnecessary network overhead. This work addresses these limitations by introducing a probabilistic trust model that considers relevant factors affecting network performance, and seeks to account for those while deciding how data is routed.

\section{Background}
\subsection{AODV Protocol}
AODV \cite{aodv} is a reactive routing protocol that establishes routes only when needed, making it suitable to reduce overhead for dynamic networks. It utilizes three main message types:
\begin{itemize}
    \item Route Request (RREQ): Broadcast by source nodes to discover routes
    \item Route Reply (RREP): Sent by destination or intermediate nodes with valid routes
    \item Route Error (RERR): Used to notify nodes of link failures
\end{itemize}

The protocol maintains sequence numbers to ensure loop freedom and uses routing tables at each node to store:
\begin{itemize}
    \item Destination address
    \item Next hop
    \item Number of hops
    \item Destination sequence number
    \item Active neighbors
    \item Expiration time
\end{itemize}

This information enables each intermediate node to maintain efficient routing by storing the respective next hops to previously encountered sources and destinations, along with the number of hops required to reach them.

\subsection{AOMDV Protocol}
AOMDV \cite{aomdv} extends AODV by computing multiple loop-free and node-disjoint or link-disjoint paths during route discovery. In this enhanced protocol, an intermediate node can store multiple next hops to a destination, along with the respective hop counts to reach the destination through that node.

\section{Proposed Trust-Based Enhancement}
\subsection{Trust Model}
The proposed enhancement introduces a probabilistic trust model that enhances the performance of existing protocols:

\subsubsection{Link Trust}
The trust value of a link represents the historical reliability of the corresponding node. This trust value is modeled using a Beta distribution, which is particularly well-suited for representing the probability of a binary event (successful or failed packet delivery):

\begin{equation}
    \label{eq:beta-dist}
    p_{A \rightarrow B} \sim \text{Beta}(\alpha, \beta)
\end{equation}

where $p_{A \rightarrow B}$ represents the probability of node A successfully receiving an acknowledgment (ACK) from node B after sending a packet. The expected value of this trust value is:

\begin{equation}
    \label{eq:expected-trust}
    E[p_{A \rightarrow B}] = \frac{\alpha}{\alpha + \beta}
\end{equation}

\begin{proof}
The Beta distribution is parameterized by two positive shape parameters, $\alpha$ and $\beta$. The probability density function of the Beta distribution is:

\[f(x; \alpha, \beta) = \frac{\Gamma(\alpha + \beta)}{\Gamma(\alpha)\Gamma(\beta)} x^{\alpha-1}(1-x)^{\beta-1}\]

where $\Gamma$ is the Gamma function.

The expected value of a random variable $X$ following a Beta distribution is given by:

\[E[X] = \frac{\alpha}{\alpha + \beta}\]

Substituting $X = p_{A \rightarrow B}$, we obtain the desired expression for the expected trust value as shown in \cref{eq:expected-trust}.
\end{proof}

To update trust between nodes A and B, a Beta prior combined with a Bernoulli likelihood is used. \Cref{alg:trust-update} modifies the Beta parameters $\alpha$ and $\beta$ based on A's experience of receiving acknowledgments (ACK) from B.

\begin{algorithm}
    \caption{Trust Update Algorithm for A to B}
    \label{alg:trust-update}
    \begin{algorithmic}[1]
        \State Send message to node B
        \If{ACK received from B}
            \State $\alpha \leftarrow \alpha + 1$
        \Else
            \State $\beta \leftarrow \beta + 1$
        \EndIf
    \end{algorithmic}
\end{algorithm}

To account for the "inertia" of the trust values introduced through increasing values of $\alpha$ and $\beta$, and adapt to the ad-hoc nature of the network, $\alpha$ and $\beta$ are determined through only the latest $N$ observations of a queue, which maintains the reception of ACKs for sent packets. This approach allows the trust model to adapt to changes in network conditions while maintaining a stable history.

\subsubsection{Battery Level Consideration}
Battery levels are periodically broadcasted by nodes and are utilized by their neighbors as a reliability metric. The future battery level of a node is estimated based on its current level and estimated discharge rate as:

\begin{equation}
    \label{eq:battery-level}
    B_{t_2} = B_{t_1} + \frac{B_{t_1} - B_{t_0}}{t_1 - t_0} \cdot (t_2 - t_1)
\end{equation}

where $t_2 > t_1 > t_0$ and $B_{t_1}$ and $B_{t_0}$ are the battery levels at times $t_1$ and $t_0$, respectively, and are the closest observations in time to $t_2$.

The probability influence of $B_{t_2}$, denoted as $f(B_{t_2})$, can be modeled as a relevant increasing function, for e.g. $f(x) = 1 - e^{-\gamma x}$, with $\gamma$ as a chosen parameter.

\subsubsection{Path Coupling}
To account for interference between neighboring nodes, an Availability Index $A_i(t)$ is defined for each node $i$, separately stored by each neighbor. This index reflects the node's recent transmission status, as detected by the prospective sender:

\begin{equation}
    \label{eq:availability-index}
    A_i(t + \Delta t) = \min\left(1, A_i(t) + r\Delta t - \mathbb{1}_{\{i \text{ transmitting}\}} \cdot d\right)
\end{equation}

where $r$ represents the regeneration rate when idle, $d$ is the degradation factor during transmission, and $\mathbb{1}_{\{\cdot\}}$ is the indicator function.

\subsubsection{Higher Hop Path Consideration}
\begin{figure}[ht]
    \centering
    \includegraphics[width=0.6\linewidth]{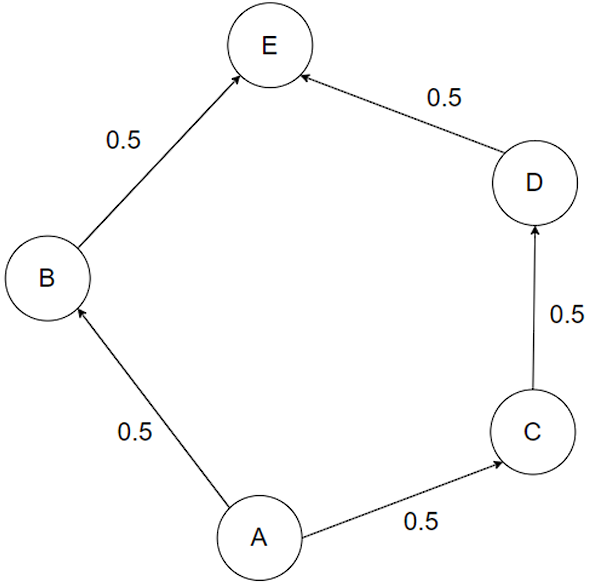}
    \caption{Illustration of hop count consideration in path selection}
    \label{fig:hop-count}
\end{figure}

As illustrated in \cref{fig:hop-count}, to account for paths with different hop counts, link trust values are adjusted by being divided by the number of hops to reach the destination through that node:

\begin{equation}
    \label{eq:hop-adjustment}
    p' = \frac{p}{\text{num\_hops}}
\end{equation}

This adjustment assumes the trust value of all links along the path is similar to that of the first hop, and that delay is proportional to the number of hops.

\subsubsection{Final probability}
The final link probability incorporates all previously mentioned factors, considering previous outcomes, battery levels, path coupling, and hop count:

\begin{equation}
    \label{eq:final-probability}
    p_{A \rightarrow B, t} = \frac{\alpha}{N} \cdot f(B_{t}) \cdot A_i(t) \cdot \frac{1}{\text{num\_hops}}
\end{equation}

\subsection{Expected Delay Calculation}
The expected delay for a path is computed as:
\begin{equation}
    \label{eq:delay-recursive}
    E[\text{delay}] = p \cdot 2t + (1 - p) \cdot (2t + E[\text{delay}])
\end{equation}
\begin{equation}
    \label{eq:delay-final}
    \therefore E[\text{delay}] = \frac{2t}{p}
\end{equation}

where $t$ represents the average transmission time and $p$ is the successful transmission probability. A new data packet will be transmitted only after receiving the ACK for the previous one.

From \cref{eq:delay-final}, it can be deduced that:
\begin{equation}
    \label{eq:delay-proportion}
    E[\text{delay}] \propto \frac{1}{p}
\end{equation}

\subsection{Optimal Data Distribution}
\begin{figure}[ht]
    \centering
    \includegraphics[width=0.5\linewidth]{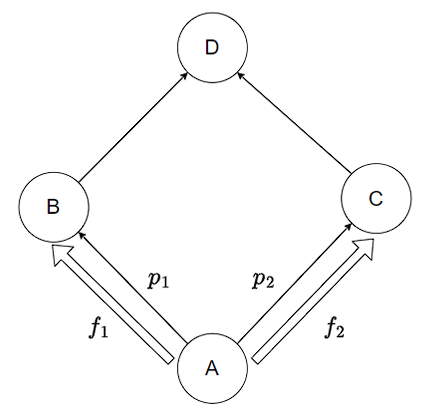}
    \caption{Optimization strategy for data distribution across multiple paths}
    \label{fig:optim-ratio}
\end{figure}

In the scenario shown in \cref{fig:optim-ratio}, the data distribution across the two paths is chosen to minimize the expected delay of completely offloading transmitted data from A to B and C. For data distribution ratio $f_1 : (1-f_1)$, from \cref{eq:delay-proportion} and assuming Delay $\propto$ amount of data being transmitted:

\begin{itemize}
    \item Delay for transmission along path 1 $\propto \frac{f_1}{p_1}$
    \item Delay for transmission along path 2 $\propto \frac{1 - f_1}{p_2}$
\end{itemize}

To minimize delay for complete transmission, the following needs to be minimized:
\[\max \left(\frac{f_1}{p_1}, \frac{1 - f_1}{p_2}\right)\]

Therefore:
\begin{equation}
    \label{eq:optimal-ratio}
    f_1 = \frac{p_1}{p_1 + p_2}
\end{equation}

where $p_1$ and $p_2$ are the link-trust values of the first and second paths, respectively.

This result generalizes to $n$ paths:
\begin{equation}
    \label{eq:general-ratio}
    f_i = \frac{p_i}{\sum_{j=1}^n p_j}
\end{equation}

where $n$ is the number of neighbors of the transmitting node.

\section{Conclusions and Future Work}
\begin{figure}[ht]
    \centering
    \includegraphics[width=0.9\linewidth]{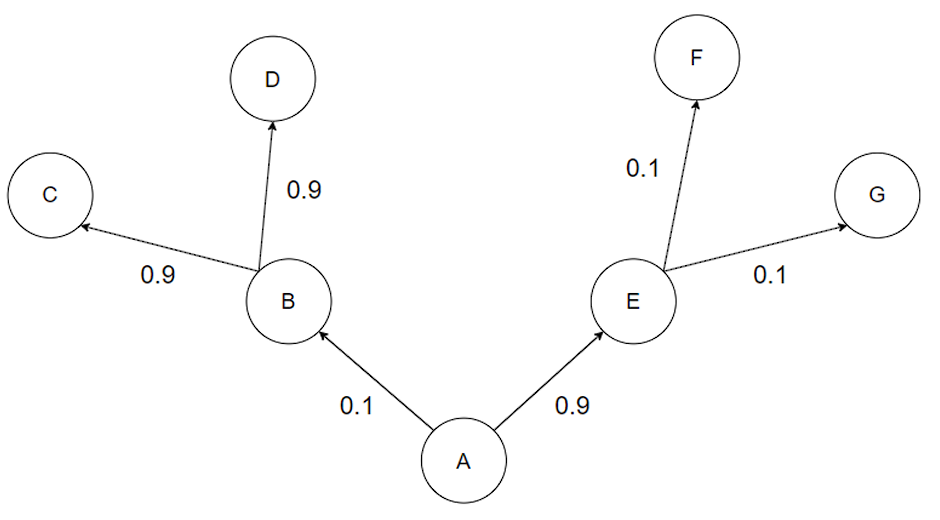}
    \caption{Key considerations for future protocol enhancements}
    \label{fig:future-work}
\end{figure}

This paper presented a probabilistic trust-based enhancement to AOMDV that incorporates factors such as past behavior, battery levels, and node coupling to distribute data optimally to reduce delay, while simultaneously transmitting it through multiple paths.
The next step involves simulating the method and analyzing observations.
Future improvements and research directions include:
\begin{itemize}
    \item Improving visibility beyond immediate neighbors to incorporate trust values of more links along the path to the destination, while minimizing overhead. This arises from the need to address the scenario depicted in \cref{fig:future-work} where data might be routed locally optimally but globally unoptimally. 
    \item Including queue lengths of recipient node as another parameter while calculating link trust, decreasing it as detected queue size increases.
    \item Dynamically adapting parameters, including queue length $N$ and $\gamma$ in the battery function.
    \item Modifying AOMDV to store more paths to the destination at intermediate nodes.
    \item Optimally consolidating probabilities from individual metrics.
\end{itemize}

\section{Acknowledgment}
This paper was developed in large part through discussions in the course EE798C (Advanced Networking), taught by Professor Yatindra Nath Singh at IIT Kanpur. I am grateful for his suggestions and invaluable feedback on my work during the course.

\bibliographystyle{IEEEtran}

\end{document}